\documentclass{amsart} 
\usepackage{graphicx}
\usepackage{amssymb, amsmath, sidecap}
\usepackage{amsfonts}
\usepackage{amssymb}
\usepackage{float}
\usepackage[version=3]{mhchem}
\usepackage{endnotes}

  \newcommand{\p}{\varphi}

\def\f{\frac}
 
\def\bt{\begin{thm}}
\def\et{\end{thm}}
\def\bl{\begin{lem}}
\def\el{\end{lem}}
\def\bd{\begin{defi}}
\def\ed{\end{defi}}
\def\bc{\begin{cor}}
\def\ec{\end{cor}}
\def\bp{\begin{proof}}
\def\ep{\end{proof}}
\def\br{\begin{rem}}
\def\er{\end{rem}}
\def\Le{\text{\rm Le }}
\def\Len{\text{\rm Le}}
\def\Pr{\text{Pr}}
\def\p{\partial}
\def\Re{\mathfrak{Re}}

\def\bcon{\begin{conclusion}}
\def\econ{\end{conclusion}}

\newtheorem{thm}{Theorem}[section]
\newtheorem{lem}{Lemma}[section]
\newtheorem{defi}{Definition}[section]

\newtheorem{rem}{Remark}[section]
\newtheorem{cor}{Corollary}[section]

\newtheorem{conclusion}{Physical Conclusion}[section]

\numberwithin{equation}{section}
\numberwithin{figure}{section}

\begin{document}

\title{Tropical Atmospheric Circulations with Humidity Effects}
\author[C. Hsia]{Chun-Hsiung Hsia}
\address[CH]{Department of Mathematics, National Taiwan University, Taipei 10617, Taiwan }
\email{willhsia@math.ntu.edu.tw}

\author[C. S. Lin]{Chang-Shou Lin}
\address[CL]{Department of Mathematics,
National Taiwan University, Taipei 10617, Taiwan}
\email{cslin@math.ntu.edu.tw}

\author[T. Ma]{Tian Ma}
\address[TM]{Department of Mathematics, Sichuan University, Chengdu,
 P. R. China}
 \email{matian56@sina.com}

\author[S. Wang]{Shouhong Wang}
\address[SW]{Department of Mathematics,
Indiana University, Bloomington, IN 47405, USA}
\email{showang@indiana.edu}
\thanks{The work of SW was supported in part by the
US Office of Naval Research and by the US National Science Foundation.}


\begin{abstract}
The  main objective of this article is to study the effect of the moisture on the planetary scale atmospheric circulation over the tropics. The modeling we adopt is the Boussinesq equations coupled with a diffusive equation of humidity and the humidity dependent heat source is modeled by a  linear approximation of the humidity. The rigorous mathematical analysis is carried out using the dynamic transition theory. In particular, we obtain the same types of transitions and hence the scenario of the El Ni\~no mechanism as described in \cite{MW2,MW3}. The effect of the moisture only lowers slightly the magnitude of the critical thermal Rayleigh number.

\end{abstract}
\maketitle
\tableofcontents

\section{Introduction}
This article is part of a research program to study low frequency variability of the atmospheric and oceanic flows.
As we know, typical sources of climate low frequency variability include the  wind-driven (horizontal) and thermohaline (vertical) circulations (THC) of the ocean, and the El Ni\~no Southern Oscillation (ENSO).  Their variability, independently and interactively,  may play a significant role  in climate change, past and future.
The  primary goal  of our study  is to document, through careful theoretical and numerical studies, the presence of climate low frequency variability,
to verify the robustness of this variability's characteristics to changes in model parameters, and to help explain its physical mechanisms. The thorough understanding of the variability is a challenging problem with important practical implications for geophysical efforts to quantify predictability, analyze error growth in dynamical models, and develop efficient forecast methods.

ENSO is one of the strongest interannual climate variability associated with strong atmosphere-ocean coupling, with significant impacts on global climate. ENSO consists of  warm events (El Ni\~no phase) and cold events (La Ni\~na phase) as observed by the equatorial eastern Pacific SST anomalies, which are associated with persistent weakening or strengthening in the trade winds; see among others 
\cite{BH,jin,JNG,neelin90,neelin91,SS,ZC,neelin,ghil00,penland,tribbia,
PF,LW08,samelson08}.  An interesting current debate is whether ENSO is best modeled as a stochastic or chaotic system - linear and noise-forced, or nonlinear oscillatory and unstable system  \cite{PF}?  It is obvious that a careful fundamental level examination of the problem is crucial. For this purpose, Ma and Wang \cite{MW2} initiated a study of ENSO from the dynamical transition point of view, and derived in particular a new oscillation mechanism of the ENSO. Namely, ENSO is  a self-organizing and self-excitation system, with two highly coupled oscillation processes--the oscillation between metastable El Nino and La Nina  and normal states, and the spatiotemporal oscillation of the sea-surface temperature (SST).

The main objective of this article is to address the moisture effect on the low frequency variability associated with the ENSO.  First, as our main purpose is to capture the patterns and general features of the large scale atmospheric circulation over the tropics,  it is appropriate to use the Boussinesq equations coupled with a diffusive equation of humidity. In addition, the humidity effect is also taking into consideration by treating the  heating source as  a  linear approximation of the humidity  function.

Second, although  the introduction of the humidity effect leads to substantial difficulty from the mathematical point of view, we haver shown   in Theorem 4.1  that  the humidity
 does not affect the type of dynamic transition  the system undergoes. Namely, we show that under the idealized boundary conditions, only the continuous transition  (Type-I) occurs. However, the critical thermal Rayleigh  number is  slightly smaller than that in the case without moisture factor. To see this effect of the humidity, we refer to formula (\ref{eq4.23}) which reads
 \begin{equation*}
 \begin{aligned}
 R_c = &   -\left[\frac{1}{\Len}+\left(1+\frac{\Le \alpha_{k_c}^2}{24}\right) \frac{\alpha_1 \alpha_T h^2}{\Le \alpha_q \alpha_{k_c}^2 \kappa_T}    \right] \tilde{R}  + \alpha^4_{k_c}\frac{r_0^2}{k^2} \left(\alpha^2_{k_c}+\frac{1}{r_0^2}\right)+\frac{\alpha_{k_c}^2}{r_0^2}. \\
 \end{aligned}
 \end{equation*}
Since in the expression of thermal critical Rayleigh number $R_c$, the coefficient of the humidity Rayleigh number $\tilde{R}$ is negative,  and  this indicates that the presence of humidity lowers down the critical temperature difference for  the onset of the dynamic transition. 

Third, we remark that the perturbation analysis  in \cite{MW2}  can be applied to the case here to carry out the analysis, and   we can  show that   that under the natural boundary condition,  the  underlying system with humidity effect will undergo   a mixed type transition. In addition, 
as we argued in \cite{MW2}, it is necessary to include turbulent friction terms in the model to obtain correct convections scales for the large scale tropical atmospheric circulations, leading   in particular  to the right critical temperature gradient and the length scale for the Walker circulation. 

Finally, based on these theoretical results, it is easy then to conclude the same mechanism for ENSO as proposed  in Ma and Wang \cite{MW2. MW3}. Namely, ENSO is  a 
self-organizing and self-excitation system, with  two highly coupled oscillatory processes: 1) the oscillation between the two metastable  warm (El Nino phase)  and cold events (La Nina phase), and 2) the spatiotemporal oscillation of the  sea surface temperature (SST) field. The interplay between these two processes gives rises the climate variability associated with the ENSO, leads to both the random and deterministic features of the ENSO, and defines a new natural feedback mechanism, which drives the sporadic oscillation of the ENSO.  With the deterministic model considered in this article, the randomness is closely related to the uncertainty/fluctuations of the initial data between the narrow basins of attractions of the corresponding metastable events, and the deterministic feature is represented by a deterministic coupled atmospheric and oceanic model predicting the basins of attraction and the sea-surface temperature (SST). 

It is worth mentioning that as practiced by many  in the field, the randomness associated with the ENSO may be incorporated by introducing the noise directly in the model. In addition, from the predictability and  prediction point of view, it crucial to  capture more detailed information on the delay feedback mechanism of the SST. For this purpose, the study of explicit multi-scale coupling mechanism to the ocean is inevitable. 

This article is organized as follows. Section 2 gives the objective Boussinesq model with humidity. The eigenvalue problem is analyzed in Section 3. The transition theorem is stated and proved in Section 4. In section 5, the turbulence friction factors are considered and the corresponding critical temperature difference and the wave numbers of Walker circulation are checked.

\section{Model for Atmospheric Motion  with Humidity}

\subsection{Atmospheric circulation model}

The hydrodynamical equations governing the atmospheric circulation is the Navier-Stokes equations with the Coriolis force generated by the earth's rotation,
coupled with the first law of thermodynamics.

Let $(\varphi,\theta, r)$ be the spheric coordinates, where $\varphi$ represents the longitude, $\theta$ the latitude, and $r$ the radial coordinate. The unknown functions include  the  velocity field  $u=(u_{\varphi},u_{\theta},u_r)$, the temperature function $T$, the humidity function $q$, the pressure $p$ and the density function $\rho$.
Then the equations governing the motion and states of the atmosphere consist of the momentum equation, the continuity equation, the first law of thermodynamic, the diffusion equation for humidity, and the equation of state (for ideal gas) which read
\begin{align}
& \rho \left[ \frac{\partial u}{\partial t}+ \nabla_u u + 2 {\vec \Omega} \times u\right] + \nabla p + \vec k \rho g = \mu\triangle u, \label{10.1}\\
& \frac{\partial \rho}{\partial t} + \text{div}(\rho u)=0,\label{10.2} \\
& \rho c_v \left[ \f{\p  T}{\p t} +  u \cdot \nabla T \right]
+ p  \text{div } u=Q + \kappa_T \triangle T, \label{10.3}\\
&  \rho \left[\f{\p  q}{\p t} + u  \cdot \nabla q\right] =S + \kappa_q \triangle q, \label{10.4}\\
& p= R\rho T.\label{10.5}
\end{align}
Here  $0\leq  \varphi\leq 2\pi$,  $-{\pi}/{2}\leq\theta\leq\pi/{2}$,  $a< r<a+h, a$ is
the radius of the earth, $h$ is the height of the troposphere, $\Omega$ is
the earth's rotating angular velocity,
$g$ is the gravitative constant, $\mu$, $\kappa_T, \kappa_q, c_v, R$ are constants, $Q$  and $S$ are heat and humidity sources, and $k=(0, 0, 1)$. The  differential
operators used are as follows:

1. The gradient and divergence operators are given by:
\begin{align*}
& \nabla = \left( \frac{1}{r\cos\theta}\frac{\partial}{\partial\varphi}, \frac{1}{r}\frac{\partial}{\partial\theta}, \frac{\partial}{\partial r} \right), \\
&\text{div}\ u =\frac{1}{r^2}\frac{\partial}{\partial
r}(r^2u_r)+\frac{1}{r\cos\theta}\frac{\partial (u_{\theta}\cos\theta
)}{\partial\theta}+\frac{1}{r\cos\theta}\frac{\partial
u_{\varphi}}{\partial\varphi};
\end{align*}

2. In the spherical geometry, although the Laplacian for  a scalar is different from the Laplacian for a vectorial function, we use the same notation $\triangle$ for both of them:
\begin{align*}
& \triangle u = \Big(
 \Delta u_{\varphi}+\frac{2}{r^2\cos\theta}\frac{\partial
u_r}{\partial\varphi}+\frac{2\sin\theta}{r^2\cos^2\theta}\frac{\partial
u_{\theta}}{\partial\varphi}-\frac{u_{\varphi}}{r^2\cos^2\theta},\\
&  \qquad  \Delta
u_{\theta}+\frac{2}{r^2}\frac{\partial
u_r}{\partial\theta}-\frac{u_{\theta}}{r^2\cos^2\theta}-\frac{2\sin\theta}{r^2\cos^2\theta}
\frac{\partial u_{\varphi}}{\partial\varphi},\\
& \qquad  \Delta
u_r-\frac{2u_r}{r^2}-\frac{2}{r^2\cos\theta}\frac{\partial
(u_{\theta}\cos\theta
)}{\partial\theta}-\frac{2}{r^2\cos\theta}\frac{\partial
u_{\varphi}}{\partial\varphi}\Big),\\
&\Delta f
=\frac{1}{r^2\cos\theta}\frac{\partial f}{\partial\theta}(\cos\theta\frac{\partial}{\partial\theta}
)+\frac{1}{r^2\cos^2\theta}\frac{\partial^2 f}{\partial\varphi^2}+\frac{1}{r^2}\frac{\partial f}{\partial
r}(r^2\frac{\partial}{\partial r}).
\end{align*}

3. The convection terms are given by
\begin{align*}
& \nabla_u u =\Big( u \cdot \nabla u_{\varphi}+\frac{u_{\varphi}u_r}{r}-\frac{u_{\varphi}u_{\theta}}{r}\tan \theta, \\
  &\qquad u \cdot \nabla u_{\theta}+\frac{u_{\theta}u_r}{r}+\frac{u^2_{\varphi}}{r}\tan \theta,
              u \cdot \nabla u_r-\frac{u^2_{\varphi}+u^2_{\theta}}{r} \Big),
              \end{align*}

4. The Coriolis term $2 \vec \Omega \times u$  is given by
$$             2 \vec \Omega \times u =
2 \Omega ( \cos \theta u_r - \sin \theta u_\theta, \sin \theta u_\varphi, - \cos \theta u_\varphi).$$
Here $\vec \Omega$ is the angular velocity vector of the earth, and $\Omega$ is the magnitude of the angular velocity.

The above system of equations were basically the equations used by L. F. Richardson in his pioneering work \cite{richardson}. However, they are in general too complicated to conduct theoretical analysis.
As practiced by the earlier researchers
such as J. Charney, and from
the lessons learned by the failure of Richardson's pioneering work,
one tries to be satisfied with simplified models approximating
the actual motions to a greater or less degree
instead of attempting to deal with the atmosphere in all its complexity.
By starting with models incorporating only
what are thought to be the most important of atmospheric influences,
and by gradually bringing in others, one is able to proceed
inductively and thereby
to avoid the pitfalls inevitably encountered when a great
many poorly understood factors are introduced all at once.
The simplifications are usually done by taking into consideration of some main characterizations of the large-scale atmosphere. One such characterization is the small aspect ratio between the vertical and horizontal scales, leading to hydrostatic equation replacing the vertical momentum equation. The resulting system of equation are called the primitive equations (PEs); see among others \cite{LTW92a}. The another characterization of the large scale motion is the fast rotation of the earth, leading to the celebrated quasi-geostrophic equations \cite{charney47}.

\subsection{Tropical atmospheric circulation model}

In this article, our main focus is on the formation and transitions of the general circulation patterns. For this purpose, the approximations we adopt involves the following components.

First, we often use Boussinesq assumption, where the density is treated as a constant except in the buoyancy term and in the equation of state.

Second, because the air is generally not incompressible, we do not use the equation of state for ideal gas, rather, we use the following empirical formula, which can be regarded as the linear approximation of (\ref{10.5}):
\begin{equation}
\rho =\rho_0[1 -\alpha_T  (T-T_0)  - \alpha_q (q-q_0)],\label{10.6}
\end{equation}
where $\rho_0$ is the density at $T=T_0$  and $q=q_0$,   and $\alpha_T$   and $\alpha_q$ are  the coefficients of
thermal   and humidity expansion.

Third, since the aspect ratio between the vertical scale and the horizontal scale is small,
the spheric shell the air occupies is treated as a product space $S^2_{a} \times (a, a+h)$ with the product metric 
$$
ds^2 = a^2 d \theta^2 + a^2 \sin^2 \theta d \phi^2 + dz^2.
$$

This approximation is extensively adopted in geophysical
fluid dynamics.

Fourth, the hydrodynamic equations governing the atmospheric circulation over the tropical zone are the Navier-Stokes equations coupled with the first law of thermodynamics and the diffusion equation of the humidity. These equations are restricted on the lower latitude region where the meridianal velocity component $u_{\theta}$ is zero.

Let $( \phi, z) \in M = (0, 2\pi)\times(a, a+h)$ be the coordinate where $\phi$ is the longitude, $a$ the radius of the earth, and $h$ the height of the troposphere. The unknown functions include the velocity field $u=(u_{\phi}, u_{z})$, the temperature function $T$, the humidity function $q$ and the pressure $p$. Then the equations governing the motion and states of the atmosphere read
\begin{equation}
\label{eq2.1}
\begin{aligned}
& \frac{\partial u_{\phi}}{\partial t} + (u \cdot \nabla) u_{\phi} + \frac{u_{\phi} u_{z}}{ a}=  \nu ( \Delta u_{\phi} + \frac{2}{a^2} \frac{\partial u_z}{ \partial \phi} - \frac{ u_{\phi}}{a^2}) -2 \Omega u_z - \frac{1}{ \rho_0 a} \frac{\partial p}{ \partial \phi},\\
& \frac{\partial u_z}{\partial t}+(u \cdot \nabla) u_z-\frac{u_{\phi}^2}{a}=  \nu(\Delta u_z-\frac{2}{a^2}\frac{\partial u_{\phi}}{\partial \phi} - \frac{2 u_z}{ a^2}) + 2 \Omega u_{\phi} - \frac{1}{\rho_0}\frac{\partial p}{ \partial z} \\
&  \qquad \qquad        - g (1 - \alpha_T (T-T_0)-\alpha_q(q- q_0)), \\
& \frac{\partial T}{\partial t} + (u \cdot \nabla) T = \kappa_{T}\Delta T + Q,\\
& \frac{\partial q}{ \partial t} +(u \cdot \nabla) q = \kappa_{q}\Delta q + H,\\
& \frac{1}{a} \frac{\partial u_{\phi}}{\partial \phi} + \frac{\partial u_z}{ \partial z} =0,
\end{aligned}
\end{equation}
where $\Omega$ is the earth's rotating angular velocity, $g$ the gravitational constant, $\rho_0$ the density of air at $T=T_0$, $\nu$ the viscosity, $\alpha_T$ and $\alpha_q$ are the coefficients  of thermal and humidity expansion, $\kappa_T$ and $\kappa_q$ the thermal and humidity diffusion coefficients, and $Q$, $H$ the heat and humidity sources. However, the differential operators used in (\ref{eq2.1}) are as follows
\begin{align*}
(u \cdot \nabla)& = \frac{u_{\phi}}{a} \frac{\partial}{ \partial \phi} + u_z \frac{\partial}{\partial z},\\
  \Delta f & = \frac{1}{a^2} \frac{\partial^2}{ \partial \phi^2}f + \frac{\partial^2}{\partial z^2}f, \\
 \Delta u & = \Delta (u_{\phi}e_{\phi} + u_z e_z)\\
                 & =(\Delta u_\phi + \frac{2}{a^2} \frac{\partial u_z}{\partial \phi} -\frac{ u_\phi}{ a^2 } )e_\theta + (\Delta u_z -\frac{2}{a^2} \frac{\partial u_\phi}{\partial \phi} - \frac{2u_z}{a^2} )e_z,
   \end{align*}
   where $\Delta f$ is the Laplacian for scalar functions, and $\Delta u$ is the Laplace-Beltrami operator on $(0, 2\pi) \times (a, a+h)$ with product metric.
Based on the thermodynamics, the heat source $Q$ is a function of the humidity $q$. For simplicity, we take the linear approximation
\begin{equation}
\label{eq2.2}
Q=\alpha_0 + \alpha_1 q,
\end{equation}
and the humidity source is taken as zero
\begin{equation}
\label{eq2.3}
H=0.
\end{equation}
The boundary conditions are periodic in $\phi$-direction
\begin{equation}
\label{eq2.4}
(u, T, q)(\phi + 2 \pi, z) = (u, T, q)(\phi, z),
 \end{equation}
and  free-slip on the earth surface $z=a$ and the tropopause $z=a+h$,
\begin{equation}
\label{eq2.5}
\begin{aligned}
& u_z =0, &&\frac{\partial u_{\phi}}{\partial z} =0, &&&T=T_0, &&&&q=q_0, &&&&&\text{at } z=a,\\
& u_z=0, &&\frac{u_{\phi}}{\partial z} =0, &&&T=T_1, &&&&q=0, &&&&&\text{at } z=a+ h,
\end{aligned}
  \end{equation}
where $T_0$, $T_1$, $q_0$ and $q_1$ are constants satisfying
\begin{equation}
\label{eq2.6}
T_0>T_1,\,\,\, q_0 > 0.
\end{equation}
The problem (\ref{eq2.1}) - (\ref{eq2.5}) possesses a steady state solution
\begin{equation}
\label{eqq2.13}
\begin{aligned}
& \tilde{u} =0,\\
&\tilde{T} = \frac{\gamma_1}{ 6 \kappa_T} z^3 - \frac{\gamma_0}{ 2 \kappa_T} z^2 + c_1z + c_0, \\
&\tilde{q} = -\frac{q_0}{h} z +\frac{q_0}{h} (h+a), \\
&\tilde{p}=-\int_0^z \rho_0 g \Big(  1-\alpha_T(\tilde{T}-T_0) - \alpha_q (\tilde{q} - q_0) \Big) dz,
\end{aligned}
\end{equation}
where
\begin{equation*}
\begin{aligned}
& \gamma_0 = \alpha_0+ \alpha_1 q_0 (h+a)/h, \,\,\, \gamma_1 = \alpha_1 q_0/ h,\\
& c_0=\frac{a+h}{h}\Big( -\frac{\gamma_1}{6\kappa_T} a^3 + \frac{\gamma_0}{2\kappa_T} a^2 + T_0\Big) - \frac{a}{h}\Big(- \frac{\gamma_1}{6 \kappa_T}(a+h)^3+\frac{\gamma_0}{2\kappa_T}(a+h)^2 +T_1 \Big)\\
& c_1=-\frac{1}{h}(T_0 -T_1)-\frac{1}{\kappa_T}\Big(  \frac{\gamma_1}{6}(h^2 + 3 h a +3 a^2) - \frac{\gamma_0}{2}(h+2 a) \Big)
\end{aligned}
\end{equation*}

To obtain the nondimensional form, let
\begin{equation*}
\begin{aligned}
& x=h x', &&  a=h r_0,\\
& t=h^2 t'/\kappa_T, && u = \kappa_T u'/ h,\\
 & T=(T_0-T_1)T'+\tilde{T}, && q = q_0 q'+\tilde{q},\\
  & p=\rho_0 \nu \kappa_T p'/h^2 +\tilde{p},
\end{aligned}
\end{equation*}
and consider
$$
(x_1, x_2) = (r_0 \phi, z), \,\, (u_1, u_2) = (u_{\phi}, u_z).
$$

Omitting the primes, the equations (\ref{eq2.1}) become
\begin{equation}
\label{eq2.7}
\begin{aligned}
&\frac{\partial u_1}{ \partial t} = \Pr \Big( \Delta u_1 + \frac{2}{r_0} \frac{\partial u_2}{ \partial x_1} - \frac{1}{r_0^2} u_1 -\frac{\partial p}{\partial x_1} \Big) \\
 & \qquad - \omega u_2 - (u\cdot \nabla) u_1 - \frac{1}{r_0} u_1 u_2, \\
&\frac{\partial u_2}{ \partial t} = \Pr \Big( \Delta u_2 - \frac{2}{r_0} \frac{\partial u_1}{ \partial x_1}  -\frac{2}{r_0^2} u_2 +RT + \tilde{R} q -\frac{\partial p}{ \partial x_2} \Big)\\
& \qquad +\omega u_1-(u\cdot \nabla)u_2 + \frac{1}{r_0} u_1^2,\\
 & \frac{\partial T}{\partial t} = \Delta T -\frac{1}{T_0-T_1} \frac{d \tilde{T}(h x_2)}{d x_2}u_2 + \alpha q - (u \cdot \nabla) T,\\
 & \frac{\partial q}{\partial t} = \Le \Delta q + u_2 - (u \cdot \nabla)q,\\
 & \frac{\partial u_1}{ \partial x_1} + \frac{\partial u_2}{\partial x_2}=0
\end{aligned}
\end{equation}
where the nondimensional physical parameters are
\begin{equation}
\label{eq2.8}
\begin{aligned}
& \Pr=\frac{\nu}{\kappa_T}     && \text{the Prandtl number},\\
& R= \frac{\alpha_T g (T_0 -T_1) h^3}{ \kappa_T \nu}  && \text{the thermal Rayleigh number},\\
& \tilde{R}= \frac{\alpha_q g q_0 h^3}{\kappa_T \nu}  && \text{the humidity Rayleigh number},\\
&\Len = \frac{\kappa_q}{ \kappa_T}    && \text{the Lewis number},\\
 & \omega = \frac{2 \Omega h^2}{\kappa_T}   && \text{the earth rotation}, \\
 & \alpha = \frac{\alpha_1 q_0 h^2}{ \kappa_T (T_0 -T_1)},
\end{aligned}
\end{equation}
and
\begin{equation}
\label{eq2.9}
\begin{aligned}
-\frac{1}{T_0-T_1} \frac{d \tilde{T}(hx_2)}{d x_2} =& 1 + \frac{\gamma_0 h^2}{\kappa_T (T_0-T_1)}(x_2-r_0 - \frac{1}{2})\\
& - \frac{\alpha_1 q_0 h^2}{2 \kappa_T (T_0 -T_1)}(x_2^2 - r_0^2 -r_0 - \frac{1}{3}),
\end{aligned}
\end{equation}
where $r_0 < x_2 < r_0+1$, and $r_0$ is the nondimensional radius of the earth.  For simplicity, we take the average approximation $x_2 = r_0 + \frac{1}{2}$. In this case, (\ref{eq2.9}) is given by
\begin{equation*}
-\frac{1}{T_0 -T_1} \frac{d \tilde{T}}{d x_2} \Big|_{x_2 = r_0 + \frac{1}{2}} = 1 + \frac{\alpha_1 q_0 h^2}{ 24 \kappa_T (T_0-T_1)}.
 \end{equation*}
Thus, the equation (\ref{eq2.7})  is approximated by
\begin{equation}
\label{eq2.10}
\begin{aligned}
\frac{\partial u_1}{ \partial t} = & \Pr \Big( \Delta u_1 + \frac{2}{r_0} \frac{\partial u_2}{ \partial x_1} -\frac{1}{r_0^2} u_1 - \frac{\partial p}{\partial x_1}\Big) \\
&-\omega u_2 - (u \cdot \nabla) u_1 - \frac{1}{r_0}u_1 u_2,\\
\frac{u_2}{\partial t} =& \Pr \Big( \Delta u_2 - \frac{2}{r_0}\frac{\partial u_1}{ \partial x_1} - \frac{2}{r_0^2}u_2 + RT+\tilde{R}q - \frac{\partial p}{\partial x_2}  \Big) \\
& + \omega u_1 - (u \cdot \nabla) u_2 + \frac{1}{r_0} u_1^2,\\
\frac{\partial T}{\partial t}=& \Delta T + \gamma u_2 + \alpha q - (u\cdot \nabla) T,\\
\frac{\partial q}{\partial t} = & \Le \Delta q + u_2 - (u \cdot \nabla)q,\\
\text{\it div }  u & =0,
\end{aligned}
\end{equation}
where
\begin{equation}
\label{eq2.11}
\gamma = 1 + \frac{\alpha}{24}.
\end{equation}
 The domain is $M = [0, 2 \pi r_0 ] \times (r_0, r_0+ 1)$, and the boundary conditions are given by
 \begin{equation}
 \label{eq2.12}
 \begin{aligned}
 & (u, T, q) (x_1 + 2 \pi r_0, x_2)=(u, T, q)(x_1, x_2),\\
 & u_2 =0, \, \frac{\partial u_1}{\partial x_2} =0, \, T=0, \, q=0, \, \text{at }
x_2= r_0,\, r_0+1.
 \end{aligned}
 \end{equation}
 For the problem (\ref{eq2.10})-(\ref{eq2.12}), we set the spaces
 \begin{equation}
 \label{eq2.13}
 \begin{aligned}
 & H= \Big\{ (u, T, q) \in L^2(M, \mathbb R^4) \Big| \text{div } u =0, \,\, u_2 = 0 \text{ at } r_0, r_0+1 \Big \}, \\
 & H_1 = \Big \{ (u, T, q) \in H^2 (M, \mathbb R^4) \cap H \Big| (u, T, q) \text{ satisfies } (\ref{eq2.12}) \Big\}.
 \end{aligned}
 \end{equation}
 \section{Eigenvalue Problem and Principle of Exchange of Stability}
 \subsection{Eigenvalue Problem}
 To study the transition of (\ref{eq2.10})-(\ref{eq2.12}) from the basic state, we need to consider the following eigenvalue problem
 \begin{equation}
 \label{eq3.1}
 \begin{aligned}
 & \Pr\Big( \Delta u_1 + \frac{2}{r_0} \frac{\partial u_2}{ \partial x_1} - \frac{1}{r_0^2} u_1 - \frac{\partial p}{\partial x_1}  \Big) - \omega u_2 = \beta u_1,\\
 & \Pr \Big( \Delta u_2 - \frac{2}{r_0}\frac{\partial u_1}{\partial x_1}  - \frac{2}{r_0^2} u_2 + RT +\tilde{R} q - \frac{ \partial p}{\partial x_2} \Big)+ \omega u_1 = \beta u_2, \\
 & \Delta T + \gamma u_2+  \alpha q = \beta T,\\
 & \Le \Delta q + u_2 = \beta q,\\
 & \frac{\partial u_1}{ \partial x_1} + \frac{\partial u_2}{\partial x_2} = 0,
 \end{aligned}
 \end{equation}
 supplemented with the boundary conditions (\ref{eq2.12}).
 By the boundary conditions (\ref{eq2.10}), the eigenvalues of (\ref{eq3.1})  can be solved by separation of variables as follows
 \begin{equation}
 \label{eq3.2}
 \psi^1 = \left\{ \begin{aligned}
 & u^1_1 = - u_k(x_2) \sin \frac{k x_1}{r_0},\\
  & u^1_2 = v_k(x_2)\cos\frac{k x_1}{ r_0}, \\
  & T^1 = T_k (x_2)\cos \frac{k x_1}{r_0},\\
  & q^1 = q_k(x_2) \cos \frac{k x_1}{r_0},\\
  & p^1 = A_k (x_2) \cos \frac{k x_1}{r_0} - B_k (x_2)
   \sin \frac{k x_1}{r_0}
 \end{aligned}\right. \end{equation}

\begin{equation}
\label{eq3.3}
\psi^2=\left\{ \begin{aligned}
 & u^2_1 = u_k(x_2) \cos \frac{k x_1}{ r_0},\\
 & u_2^2 = v_k (x_2) \sin \frac{k x_1}{ r_0},\\
 & T^2 = T_k (x_2) \sin \frac{k x_1}{ r_0},\\
& q^2 = q_k (x_2) \sin \frac{k x_1}{ r_0},\\
& p^2= A_k(x_2) \sin \frac{k x_1}{ r_0} + B_k(x_2)\cos \frac{k x_1}{ r_0}.
\end{aligned}
\right.
 \end{equation}
 Based on the continuity equation in (\ref{eq3.1}), we have
 \begin{equation}
  \label{eq3.4}
  u_k = \frac{r_0}{ k} \frac{d v_k}{ d x_2}.
  \end{equation}
  Let
  \begin{equation}
  \label{eq3.5}
  B_k = \frac{\omega r_0}{  k  \Pr} v_k(x_2), \quad A_k = p_k + \frac{2}{r_0} v_k(x_2).
  \end{equation}
  Due to (\ref{eq2.12}), plugging (\ref{eq3.2}) or (\ref{eq3.3})
  into (\ref{eq3.1}), we obtain the following system of ordinary differential equations
  \begin{equation}
\label{eq3.6}
\begin{aligned}
& \Pr\Big( D_k^2  u_k - \frac{1}{r_0^2} u_k - \frac{k}{r_0}p_k  \Big) = \beta u_k,\\
& \Pr\Big( D^2_k v_k - \frac{2}{r_0^2}v_k + RT_k +\tilde{R}q_k -D p_k \Big)=\beta v_k,\\
& D_k^2 T_k + \gamma v_k + \alpha q_k= \beta T_k,\\
& \Le D_k^2 q_k+v_k = \beta q_k,\\
& Du_k = 0, \,\, v_k = 0, \,\, T_k =0,\,\, q_k=0,\,\, \text{ at }  x_2=r_0, r_0+1,
\end{aligned}
    \end{equation}
where
$$
D=\frac{d}{d x_2}, D_k^2 = \frac{d^2}{dx_2^2} - \frac{k^2}{r_0^2}.
$$
Plugging $v_k = \sin j \pi(x_2 - r_0)$ into (\ref{eq3.6}), we see that the eigenvalue $\beta$ satisfies the cubic equation
\begin{equation}
\label{eq3.7}
\begin{aligned}
 (\alpha_{kj}^2 + \beta)&(\Le \alpha_{kj}^2 + \beta)(\Pr \alpha_{kj}^2+\frac{\Pr}{r_0^2}+\beta)\alpha_{kj}^2
+ \frac{k^2 \Pr}{r_0^4}(\alpha_{kj}^2 + \beta)(\Le \alpha_{kj}^2 + \beta)\\
& - \frac{k^2 \Pr R}{r_0^2}(\gamma \Le \alpha_{kj}^2+\gamma \beta + \alpha) - \frac{k^2 \Pr \tilde{R}}{r_0^2}(\alpha_{kj}^2 + \beta) =0
 \end{aligned}
 \end{equation}
 where
 \begin{equation}
 \label{eq3.8}
 \alpha_{kj}^2 = j^2 \pi^2 + \frac{k^2}{r_0^2}  \quad (j \ge 1, k \ge 1).
 \end{equation}

\subsection{Principle of exchange of stability}
 The linear stability of the problem (\ref{eq2.10})-(\ref{eq2.12}) is dictated precisely by the eigenvalues of  (\ref{eq3.1}) which are determined by (\ref{eq3.7}). The expansion of (\ref{eq3.7}) is
 \begin{equation}
\label{eq4.1}
\begin{aligned}
& \beta^3+[(1+\Le + \Pr)\alpha_{kj}^2 + \frac{\Pr}{r_0^2} + \frac{k^2 \Pr}{r_0^4 \alpha_{kj}^2}]\beta^2 \\
&+ [(\Pr + \Le \Pr + \Len)\alpha_{kj}^4
+\frac{\Pr }{r_0^2}(1+\Len)(\alpha_{kj}^2+\frac{k^2}{r_0^2}) - \frac{\Pr k^2}{r_0^2 \alpha_{kj}^2}(\gamma R + \tilde{R})]\beta \\
& + \Le \alpha^4_{kj}(\Pr \alpha^2_{kj}+\frac{\Pr}{r_0^2})+\frac{\Pr k^2}{r_0^2 \alpha^2_{kj}}(-\alpha^2_{kj}\tilde{R}-\alpha R - \gamma \Le \alpha^2_{kj} R+\frac{\Le \alpha_{kj}^4}{r_0^2})\\
& =0.
\end{aligned}
\end{equation}
We see that $\beta = 0$ is a solution of (\ref{eq4.1}) if and only if
\begin{equation}
\label{eq4.2}
\Le \alpha^4_{kj}(\Pr \alpha^2_{kj}+\frac{\Pr}{r_0^2})+\frac{\Pr k^2}{r_0^2 \alpha^2_{kj}}(-\alpha^2_{kj}\tilde{R}-\alpha R - \gamma \Le \alpha^2_{kj} R+\frac{\Le \alpha_{kj}^4}{r_0^2} )\\
=0.
\end{equation}
Inferring from (\ref{eq2.8}) and (\ref{eq2.11}),
\begin{equation}
\label{eq4.3}
\begin{aligned}
& \alpha R =\frac{\alpha_1 \alpha_{T}g q_0 h^5}{\kappa_T^2 \nu} = \frac{\alpha_1 \alpha_{T}h^2}{\alpha_q \kappa_{T}}\tilde{R},\\
& \gamma R = R + \frac{\alpha_1 \alpha_{T}g q_0 h^5}{24 \kappa_T^2 \nu} = R + \frac{\alpha_1 \alpha_{T}h^2}{24 \alpha_q \kappa_{T}}\tilde{R}.
\end{aligned}
\end{equation}
Hence, we rewrite (\ref{eq4.2}) as
\begin{equation}
\label{eq4.4}
R=-\Big(\frac{1}{\Len}+(1+\frac{\Le \alpha_{kj}^2}{24}) \frac{\alpha_1 \alpha_T h^2}{\Le \alpha_q \alpha_{kj}^2 \kappa_T}    \Big) \tilde{R}  + \alpha^4_{kj}\frac{r_0^2}{k^2} (\alpha^2_{kj}+\frac{1}{r_0^2})+\frac{\alpha_{kj}^2}{r_0^2}.
\end{equation}

We define the critical Rayleigh number $R_c$ as
\begin{equation}
\label{eq4.5}
\begin{aligned}
R_c = & \min_{k,j \ge 1} \Big[ -\Big(\frac{1}{\Len}+(1+\frac{\Le \alpha_{kj}^2}{24}) \frac{\alpha_1 \alpha_T h^2}{\Le \alpha_q \alpha_{kj}^2 \kappa_T}    \Big) \tilde{R}  + \alpha^4_{kj}\frac{r_0^2}{k^2} (\alpha^2_{kj}+\frac{1}{r_0^2})+\frac{\alpha_{kj}^2}{r_0^2}\Big]\\
=& \min_{k \ge 1} \Big[ -\Big(\frac{1}{\Len}+(1+\frac{\Le \alpha_{k1}^2}{24}) \frac{\alpha_1 \alpha_T h^2}{\Le \alpha_q \alpha_{k1}^2 \kappa_T}    \Big) \tilde{R}  + \alpha^4_{k1}\frac{r_0^2}{k^2} (\alpha^2_{k1}+\frac{1}{r_0^2})+\frac{\alpha_{k1}^2}{r_0^2}\Big],
\end{aligned}
\end{equation}
where
$$
\alpha_{k1}^2= \pi^2 + \frac{k^2}{r_0^2}.
$$
Based on (\ref{eq4.1}) - (\ref{eq4.4}), by Definition (\ref{eq4.5}), we shall prove  the following PES later.
\bl
\label{le4.1}
Let $\beta_{jk}$ be the eigenvalues of (\ref{eq3.1}) that satisfy (\ref{eq4.1}).  Let $k_c \ge 1$ be integer minimizing (\ref{eq4.5}), and $\beta^i_{k_c}$ ($1 \le i \le 3$) be solution of (\ref{eq4.1}) with $(k, j)=(k_c, 1)$, and
$$
\Re \beta^3_{k_c} \le \Re \beta^2_{k_c} \le \Re \beta^1_{k_c}.
$$
Then $\beta^1_{k_c}$ must be real near $R=R_c$, and
\begin{equation*}
\begin{aligned}
& \beta^1_{k_c}(R)\left\{ \begin{aligned}
& < 0,  && R<R_c,\\
& =0,   && R=R_c,\\
& > 0,   && R>R_c,
\end{aligned}
\right.\\
& \Re \beta^j_{k_c}(R_c) < 0,  \quad \text{for } j=2,3.
\end{aligned}
\end{equation*}
Moreover,  we have
$$
\beta_{jk} (R_c) < 0    \quad \text{for }  \beta_{jk}  \text{ being real.}
 $$
\el
\br
{\rm
The value $R_c$ defined by (\ref{eq4.5}) is called the critical Rayleigh number of real PES, which provides the critical temperature difference $\Delta T_c = T_0 - T_1$ by $R=R_c$, i.e.,
$$
\frac{\alpha_T g h^3 \Delta T_c}{\kappa_T \nu} = R_c.
$$
Equivalently, we have
\begin{equation}
\label{eq4.6}
\begin{aligned}
\Delta T_c = &\frac{\kappa_T \nu}{ \alpha_T g h^3} \Big[ -\Big(\frac{1}{\Len}+(1+\frac{\Le \alpha_{k_c1}^2}{24}) \frac{\alpha_1 \alpha_T h^2}{\Le \alpha_q \alpha_{k_c1}^2 \kappa_T}    \Big) \tilde{R} \\
 &+ \alpha^4_{k_c1}\frac{r_0^2}{k_c^2} (\alpha^2_{k_c1}+\frac{1}{r_0^2})+\frac{\alpha_{k_c1}^2}{r_0^2}\Big],
\end{aligned}
\end{equation}
where $k_c \ge 1$ is the integer which satisfies (\ref{eq4.5}).
}
\er
Next, we consider the PES for the complex eigenvalues of (\ref{eq3.1}).
Let $\beta = i\rho_0 $ ($\rho_0 \ne 0$) be a zero of (\ref{eq4.1}).  Then
\begin{equation*}
\begin{aligned}
\rho_0^2 =& (\Pr + \Le \Pr + \Len)\alpha_{kj}^4
+\frac{\Pr}{r_0^2}(1+\Len)(\alpha_{kj}^2+\frac{k^2}{r_0^2}) - \frac{\Pr k^2}{r_0^2 \alpha_{kj}^2}(\gamma R + \tilde{R}), \\
\rho_0^2 =& \Big[  \Le \alpha^4_{kj}(\Pr \alpha^2_{kj}+\frac{\Pr}{r_0^2})+\frac{\Pr k^2}{r_0^2 \alpha^2_{kj}}(-\alpha^2_{kj}\tilde{R}-\alpha R - \gamma \Le \alpha^2_{kj} R+ \frac{\Le \alpha_{kj}^4}{r_0^2})     \Big]  \Big/ \\
&\Big[ (1+\Len + \Pr)\alpha_{kj}^2 + \frac{\Pr}{r_0^2}  \Big]
\end{aligned}
\end{equation*}
Hence, Equation (\ref{eq4.1}) has a pair of purely imaginary solution $\pm i\rho_0$ if and only if the following condition holds true
\begin{equation}
\label{eq4.7}
\begin{aligned}
&\Big[ (1+\Len + \Pr)\alpha_{kj}^2 + \frac{\Pr}{r_0^2}  \Big] \Big[(\Pr +\Len  +\Le \Pr )\alpha_{kj}^4 \\
& +\frac{\Pr}{r_0^2}(1+\Len)(\alpha_{kj}^2+\frac{k^2 }{r_0^2})   - \frac{\Pr k^2}{r_0^2 \alpha_{kj}^2}(\gamma R + \tilde{R})\Big] \\
& = \Big[  \Le \alpha^4_{kj}(\Pr \alpha^2_{kj}+\frac{2\Pr}{r_0^2})+\frac{\Pr k^2}{r_0^2 \alpha^2_{kj}}(-\alpha^2_{kj}\tilde{R}-\alpha R - \gamma \Le \alpha^2_{kj} R + \frac{\Le \alpha^4_{kj}}{r_0^2})     \Big] \\
&>0.
\end{aligned}
\end{equation}

It follows from (\ref{eq4.3}) and (\ref{eq4.7}) that
\begin{equation}
\label{eq4.8}
\begin{aligned}
& \Big(\frac{ \Pr}{r_0^2} +(\Pr + 1)\alpha^2_{kj}  \Big) R \\
& =\Big[ \frac{\alpha_1 \alpha_T h^2}{\alpha_q \kappa_T}   \Big( 1 - \frac{1}{24}\Big( \frac{\Pr}{r_0^2} + (\Pr + 1)\alpha^2_{k_c^*} \Big)\Big) -\Big(\frac{ \Pr}{r_0^2} + (\Pr + \Len)\alpha^2_{k_c^*}\Big)  \Big]  \tilde{R}  + \\
&+ \frac{r_0^2 \alpha^2_{kj}}{k^2 \Pr}\Big( \Len(1+\Len)\alpha_{kj}^6 + \Pr \Big[ (1+ \Len+ \Pr)\alpha_{kj}^2+\frac{\Pr}{r_0^2} \Big] \Big[  (1+\Len)\alpha^4_{kj} \\
&+\frac{(1+\Len)}{r_0^2}\alpha_{kj}^2 +\frac{k^2 \Len}{r_0^4} \Big]+\frac{\Pr k^2}{r_0^4}\Big[(1+ \Pr)\alpha_{kj}^2 + \frac{\Pr}{r_0^2}\Big]  \Big) ,	
\end{aligned}
\end{equation}
where $\alpha_{kj}^2$ is as defined in (\ref{eq3.8}).

According to (\ref{eq4.8}), we define the critical Rayleigh number for the complex PES as follows
\begin{equation}
\label{eq4.9}
\begin{aligned}
R^*_c =&  \min_{k, j \ge1} \frac{1}{\frac{ \Pr}{r_0^2} +(\Pr + 1)\alpha^2_{kj}}  \times\\
&\Big\{ \Big[ \frac{\alpha_1 \alpha_T h^2}{\alpha_q \kappa_T}   \Big( 1 - \frac{1}{24}\Big( \frac{\Pr}{r_0^2} + (\Pr + 1)\alpha^2_{k_c^*} \Big)\Big) -\Big(\frac{ \Pr}{r_0^2} + (\Pr + \Len)\alpha^2_{k_c^*}\Big)  \Big]  \tilde{R} \\
&+ \frac{r_0^2 \alpha^2_{kj}}{k^2 \Pr}\Big( \Len(1+\Len)\alpha_{kj}^6 + \Pr \Big[ (1+ \Len+ \Pr)\alpha_{kj}^2+\frac{\Pr}{r_0^2} \Big] \Big[  (1+\Len)\alpha^4_{kj} \\
&+\frac{(1+\Len)}{r_0^2}\alpha_{kj}^2 +\frac{k^2 \Len}{r_0^4} \Big]+\frac{\Pr k^2}{r_0^4}\Big[(1+ \Pr)\alpha_{kj}^2 + \frac{\Pr}{r_0^2}\Big]  \Big) \Big\},
\end{aligned}
\end{equation}

Then we have the following complex PES.

\bl
\label{le4.2}
Let $k_c^*$, $j_c^*$ be the integers satisfying (\ref{eq4.9}), and $\beta^1_{k_c^*j_c^*}$ and $\beta^2_{k_c^*j_c^*}$ be the pair of complex eigenvalues of (\ref{eq4.1})
with $(k, j) = (k_c^*,j_c^*)$ near $R=R^*_c$. Then,
\begin{equation*}
\Re \beta^1_{k_c^*j_c^*} = \Re \beta^2_{k_c^*j_c^*} \left\{ \begin{aligned}
& <0,   && R<R_c^*,\\
& =0,   && R=R_c^*,\\
&>0,    && R>R_c^*.
\end{aligned}
\right.
 \end{equation*}
\el
Furthermore, for all complex eigenvalues $\beta_{kj}$ of (\ref{eq3.1}), we have
$$
\Re \beta_{kj}(R^*_c) < 0,   \quad \text{for } (k,j) \ne (k_c^*,j_c^*).
$$

\subsection{Proof of Lemmas~\ref{le4.1} and ~\ref{le4.2}}
We notice that all eigenvalues of (\ref{eq3.1}) are determined by (\ref{eq4.1}) and the eigenvalue equations
\begin{equation}
\label{eq4.10}
\begin{aligned}
\Delta T +\alpha q &= \beta T, \\
 \Le \Delta q  &=\beta q,
\end{aligned}
\end{equation}
with the boundary condition (\ref{eq2.12}). It is clear that the eigenvalues of (\ref{eq4.10}) are real and negative. Hence, it suffices to prove Lemmas~\ref{le4.1} and~\ref{le4.2} for eigenvalues $\beta_{kj}$ of (\ref{eq4.1}).
We rewrite (\ref{eq4.1}) as
\begin{equation}
\label{eq4.bbb}
    \beta^3   +b_2 \beta^2 + b_1 \beta +b_0 =0,
\end{equation}
where
\begin{equation*}
\begin{aligned}
b_2=&(1+\Len + \Pr)\alpha_{kj}^2 + \frac{\Pr}{r_0^2} + \frac{k^2 \Pr}{r_0^4 \alpha_{kj}^2},\\
b_1=&  (\Pr + \Le \Pr + \Len)\alpha_{kj}^4
+\frac{\Pr }{r_0^2}(1+\Len)(\alpha_{kj}^2+\frac{k^2}{r_0^2}) - \frac{\Pr k^2}{r_0^2 \alpha_{kj}^2}(\gamma R + \tilde{R}),\\
b_0=& \Le \alpha^4_{kj}(\Pr \alpha^2_{kj}+\frac{\Pr}{r_0^2})+\frac{\Pr k^2}{r_0^2 \alpha^2_{kj}}(-\alpha^2_{kj}\tilde{R}-\alpha R - \gamma \Le \alpha^2_{kj} R+\frac{\Le \alpha_{kj}^4}{r_0^2}) .
\end{aligned}
\end{equation*}
We see that
\begin{equation}
\label{eq4.11}
b_0, b_1, b_2 >0  \quad \text{at } R=0 \text{ for all } k, j \ge 1,
\end{equation}
which implies that all real eigenvalues of (\ref{eq4.1}) are negative at $R=0$.
By the definition of $R_c$, we find
\begin{equation}
\label{eq4.12}
\begin{aligned}
& b_0 > 0,  \forall (k, j) \ne (k_c,1), \,\, 0 \le R \le R_c, \\
& b_0 \left\{ \begin{aligned}  & > 0,  \,\, R<R_c\\
& =0, \,\, R=R_c,\\
& <0, \,\, R>R_c, \\
\end{aligned}
\right.  \quad \text{for }  (k, j) = (k_c, 1).
\end{aligned}
\end{equation}
Since $b_0 = - \beta^1_{kj} \beta^2_{kj} \beta^3_{kj}$,  Lemma~\ref{le4.1} follows from (\ref{eq4.11}) and (\ref{eq4.12}).

Next, let  $\beta_{k_c^* j_c^*}$, the solution  of (\ref{eq4.1}), take the form
\begin{equation}
\label{eq4.13}
\begin{aligned}
& \beta_{k_c^*j_c^*}(R) = \lambda (R) +i \rho(R) , \\
& \lambda(R) \to 0, \,\, \rho(R) \to \rho_0, \,\, \text{ as } R \to R^*_{c}.
\end{aligned}
\end{equation}
Inserting (\ref{eq4.13}) into (\ref{eq4.bbb}), we obtain
\begin{equation}
\label{eq4.15}
\begin{aligned}
& (-3\rho^2 + b_1) \lambda + b_0 -b_2 \rho^2 + o(\lambda) = 0,\\
& -\rho^3 + \rho b_1 + 2 \rho b_2 \lambda + o(\lambda)=0.
\end{aligned}
\end{equation}
Since $\rho_0 \ne 0$, we infer from (\ref{eq4.15}) that
$$
\lambda (R) + o(\lambda) = \frac{b_0 - b_2 \rho^2}{ 3 \rho^2 - b_1}=\frac{b_0 - b_1 b_2-2b_2^2 \lambda}{2 b_1 + 6 b_2 \lambda} +o(\lambda).$$
Thus, we have
\begin{equation}
\label{eq4.16}
\begin{aligned}
\lambda(R)= \frac{b_0 - b_1 b_2-2b_2^2 \lambda}{2 b_1 + 6 b_2 \lambda} + o(\lambda).
\end{aligned}
\end{equation}
Under the condition (\ref{eq4.7}), we see $b_1>0$. It  follows from (\ref{eq4.13}) and (\ref{eq4.16}) that
\begin{equation}
\Re \beta_{k_c^*j_c^*}(R) = \lambda(R) \left\{\begin{aligned}
< 0, \quad b_0 < b_1 b_2, \\
=0,  \quad b_0 = b_1 b_2, \\
>0, \quad b_0 > b_1 b_2,
\end{aligned}\right.
 \end{equation}
for $R$ near $R^*_c$. By the definition of $R^*_c$, it is easy to see
that
\begin{equation}
b_0 \left\{ \begin{aligned}
& < b_1 b_2,  && R<R^*_c,\\
& = b_1 b_2,  && R=R^*_c,\\
& > b_1 b_2,  && R>R^*_c.
\end{aligned}
\right.
\end{equation}
This proves Lemma~\ref{le4.2}.

By Lemmas~\ref{le4.1} and ~\ref{le4.2}, we immediately obtain the following theorem which provides a criterion to determine the equilibrium and the spatiotemporal oscillation transitions.
\bt
\label{th4.1}
Let $R_c$ and $R^*_c$ be the parameters defined by (\ref{eq4.5}) and (\ref{eq4.9}) respectively. Then  the following assertions hold true.
\begin{enumerate}
\item When $R_c < R_c^*$, the first critical-crossing eigenvalue of the problem (\ref{eq3.1}) is $\beta^1_K$ given by Lemma~\ref{le4.1}, i.e.,
\begin{align}
\label{eq4.19}
& \beta^1_{k_c}(R)\left\{ \begin{aligned}
& < 0,  && \text{if } R<R_c,\\
& =0,   && \text{if } R=R_c,\\
& > 0,   && \text{if } R>R_c,
\end{aligned}
\right.\\
\label{eq4.20}
& \Re \beta(R_c) < 0,
\end{align}
for all other eigenvalues $\beta$ of  (\ref{eq3.1}).

\item When $R^*_c < R_c$, the first critical-crossing eigenvalues are the pair of complex eigenvalues $\beta^1_{k_c^*j_c^*}$ and $\beta^2_{k_c^*j_c^*}$ given by Lemma~\ref{le4.2}, namely,
\begin{align}
\label{eq4.21}
& \Re \beta^1_{k_c^*j_c^*}(R)=\Re \beta^2_{k_c^*j_c^*}(R) \left\{ \begin{aligned}
& < 0,   && \text{if } R<R^*_c,\\
& = 0,   && \text{if } R=R^*_c,\\
& > 0,   && \text{if } R>R^*_c,
\end{aligned}
\right.\\
\label{eq4.22}
& \Re \beta(R^*_c) < 0,
\end{align}
for all other eigenvalues  $\beta(R)$ of (\ref{eq3.1}).
\end{enumerate}
\et
\br
\label{rem4.2}
{\rm In the atmospheric science, the integer $j_c^*$ in (\ref{eq4.21})  is $1$.
Hence, the critical Rayleigh numbers $R_c$ and $R_c^*$
are given by
\begin{align}
\label{eq4.23} R_c = &   -\Big(\frac{1}{\Len}+(1+\frac{\Le \alpha_{k_c}^2}{24}) \frac{\alpha_1 \alpha_T h^2}{\Le \alpha_q \alpha_{k_c}^2 \kappa_T}    \Big) \tilde{R}  + \alpha^4_{k_c}\frac{r_0^2}{k^2} (\alpha^2_{k_c}+\frac{1}{r_0^2})+\frac{\alpha_{k_c}^2}{r_0^2} \\
\label{eq4.24} R^*_c =&  \frac{1}{\frac{ \Pr}{r_0^2} +(\Pr + 1)\alpha^2_{k_c^*}}  \times\\
\nonumber &\Big\{ \Big[ \frac{\alpha_1 \alpha_T h^2}{\alpha_q \kappa_T}   \Big( 1 - \frac{1}{24}\Big( \frac{\Pr}{r_0^2} + (\Pr + 1)\alpha^2_{k_c^*} \Big)\Big) -\Big(\frac{ \Pr}{r_0^2} + (\Pr + \Len)\alpha^2_{k_c^*}\Big)  \Big]  \tilde{R} \nonumber\\
\nonumber &+ \frac{r_0^2 \alpha^2_{k_c^*}}{k^2 \Pr}\Big( \Len(1+\Len)\alpha_{k_c^*}^6 + \Pr \Big[ (1+ \Len+ \Pr)\alpha_{k_c^*}^2+\frac{\Pr}{r_0^2} \Big] \Big[  (1+\Len)\alpha^4_{k_c^*} \\
\nonumber &+\frac{(1+\Len)}{r_0^2}\alpha_{k_c^*}^2 +\frac{{k_c^*}^2 \Len}{r_0^4} \Big]+\frac{\Pr {k_c^*}^2}{r_0^4}\Big[(1+ \Pr)\alpha_{k_c^*}^2 + \frac{\Pr}{r_0^2}\Big]  \Big) \Big\},
\end{align}
where
\begin{equation}
\label{eq4.25}
\alpha_{k_c}^2 = \pi^2 + \frac{k_c^2}{r_0^2} \quad   \text{and} \quad  \alpha^2_{k_c^*}=\pi^2 + \frac{{k_c^*}^2}{r_0^2}.
\end{equation}
}
\er
\section{Transition Theorem}
 Inferring from Theorem~\ref{th4.1}, the system (\ref{eq2.10}) - (\ref{eq2.12}) have a transition to equilibria at $R=R_c$ provided $R_c < R_c^*$, and have a transition to spatiotemporal oscillation at $R=R^*_c$ provided $R^*_c < R_c$.

\bt
 For the problem (\ref{eq2.10})-(\ref{eq2.12}) we have the following assertions.
\begin{enumerate}
\item[(1)] When $R<\min \{ R_c,  R^*_c \}$, the equilibrium solution
$(u, T, q)=0$ is stable in $H$.
\item[(2)] If $R_c < R^*_c$, then this problem has a continuous transition at $R=R_c$, and bifurcates from $((u,T,g),R)=(0,R_c)$ to an attractor $\Sigma_R=S^1$ on $R>R_c$ which is a cycle of steady state
solutions.
\item[(3)] As $R^*_c < R_c$, the problem has a  transition at $R=R^*_c$,
which is either of continuous type or of jump type, and it transits to a spatiotemporal oscillation solution. In particular, if the transition is continuous, then there is an attractor of $3$-dimensional homological sphere $S^3$ is bifurcated from $((u,T,q), R)=(0, R^*_c)$ on $R>R^*_c$,
which contains no steady state solutions.
\end{enumerate}
\et
\noindent {\bf Proof}  \, We shall prove this theorem by several steps.

\noindent {\it Step 1.} Let $H$  and $H_1$ be the spaces defined by (\ref{eq2.13}). We define the operators
$L_R=A+B_R: H_1 \to H$ and $G:H_1 \to H$ by
\begin{equation}
\label{eq5.2}
\begin{aligned}
& A \psi  = P\Big( \Pr \big( \Delta u_1 + \frac{2}{r_0} \frac{\partial u_2}{ \partial x_1}\big), \Pr \big( \Delta u_2 - \frac{2}{r_0} \frac{\partial u_1}{ \partial x_1}\big), \Delta T , \Le \Delta q \Big), \\
& B_{R}\psi = P\Big( -\frac{\Pr}{r_0^2}u_1 -\omega u_2, -\Pr( \frac{2}{r_0^2} u_2 -RT - \tilde{R} q )+\omega u_1, \gamma u_2+\alpha q, u_2 \Big),\\
& G(\psi)  = -P \Big( (u \cdot \nabla)u_1+\frac{1}{r_0}u_1 u_2, (u\cdot \nabla)u_2-\frac{1}{r_0}u_1^2, (u\cdot \nabla)T, (u \cdot \nabla) q)  \Big),
\end{aligned}
\end{equation}
where $P:L^2(M, \mathbb R^4) \to H$ is the Leray projection and
$$
\psi =(u, T , q) \in H_1.
$$
 Thus, the problem (\ref{eq2.10})-(\ref{eq2.12}) is expressed in form of
\begin{equation}
\label{eq5.3}
\frac{d \psi}{dt}=L_R \psi + G(\psi),
\end{equation}
and the eigenvalue problem (\ref{eq3.1}) with the condition (\ref{eq2.12}) is rewritten as
\begin{equation}
\label{eq5.4}
L_R \psi = \beta \psi.
\end{equation}

\noindent {\it Step 2.} We shall calculate the center manifold reduction for (\ref{eq5.3}) in this step. Let $\psi^1_{k_c}$ and $\psi^2_{k_c}$ be the eigenfunctions
of (\ref{eq5.4}) corresponding to $\beta^1_{k_c}(R)$, where $\beta^1_{k_c}$ is the eigenvalue of (\ref{eq5.4}) in the case of (\ref{eq4.19}). Denote the conjugate eigenfunctions of $\psi^1_{k_c}$ and $\psi^2_{k_c}$  by ${\psi^1_{k_c}}^*$ and ${\psi^2_{k_c}}^*$, i.e.,
\begin{equation}
\label{eq5.5}
L_R^* {\psi_{k_c}^i}^* = \beta^1_K {\psi_{k_c}^i}^*, \quad i=1,2.
\end{equation}
The corresponding equations of (\ref{eq5.5}) read
 \begin{equation}
 \label{eq5.6}
 \begin{aligned}
 & \Pr\Big( \Delta u_1^* + \frac{2}{r_0} \frac{\partial u_2^*}{ \partial x_1} - \frac{1}{r_0^2} u_1^* - \frac{\partial p^*}{\partial x_1}  \Big) + \omega u_2^* = \beta u_1^*,\\
 & \Pr \Big( \Delta u_2^* - \frac{2}{r_0}\frac{\partial u_1^*}{\partial x_1}  - \frac{2}{r_0^2} u_2^* - \frac{ \partial p^*}{\partial x_2} \Big)+\gamma T^* + q^*- \omega u_1^* = \beta u_2^*, \\
 & \Delta T^* + \Pr R  u_2^* = \beta T^*,\\
 & \Le \Delta q^* +\Pr \tilde{R} u_2^*+\alpha T^* = \beta q^*,\\
 & \frac{\partial u_1^*}{ \partial x_1} + \frac{\partial u_2^*}{\partial x_2} = 0,
 \end{aligned}
 \end{equation}
Express $\psi \in H_1$ as
\begin{equation}
\label{eq5.7}
\psi = x \psi^1_{k_c} + y \psi^2_{k_c} +\Phi(x,y, R),
\end{equation}
where $(x, y) \in \mathbb R^2$  and $\Phi(x, y, R)$ is the center manifold function of (\ref{eq5.3}) near $R=R_c$. Then, the reduction equations of (\ref{eq5.3}) are
\begin{equation}
\label{eq5.8}
\left\{
\begin{aligned}
\frac{dx}{dt}&= \beta^1_{k_c} x + \frac{1}{<\psi^1_{k_c},{\psi^1_{k_c}}^*>} \big<G(\psi), {\psi^1_{k_c}}^* \big>,\\
\frac{dy}{dt}&= \beta^1_{k_c} y + \frac{1}{<\psi^2_{k_c},{\psi^2_{k_c}}^*>} \big<G(\psi), {\psi^2_{k_c}}^* \big>.
\end{aligned}
\right.
\end{equation}

 \noindent {\it Step 3: Computation of $\psi^i_{k_c}$ and ${\psi^i_{k_c}}^*$ ($i=1,2$)} Based on (\ref{eq3.2}) -(\ref{eq3.8}), we can derive

\begin{equation}
 \label{eq5.9}
 \psi^1_{k_c} = \left\{ \begin{aligned}
 & u^1_1 = - u_{k_c} \cos \pi (x_2-r_0) \sin \frac{k_c x_1}{r_0},\\
  & u^1_2 = v_{k_c} \sin \pi (x_2-r_0)\cos\frac{k_c x_1}{ r_0}, \\
  & T^1_1 = T_{k_c} \sin \pi (x_2-r_0)\cos \frac{k_c x_1}{r_0},\\
  & q^1 = q_{k_c} \sin \pi (x_2-r_0) \cos \frac{k_c x_1}{r_0},
 \end{aligned}\right. \end{equation}

\begin{equation}
\label{eq5.10}
\psi^2_{k_c}=\left\{ \begin{aligned}
 & u^2_1 = u_{k_c} \cos \pi (x_2-r_0) \cos \frac{k_c x_1}{ r_0},\\
 & u_2^2 = v_{k_c} \sin \pi (x_2-r_0) \sin \frac{k_c x_1}{ r_0},\\
 & T^2 = T_{k_c} \sin \pi (x_2-r_0) \sin \frac{k_c x_1}{ r_0},\\
& q^2 = q_{k_c} \sin \pi (x_2-r_0) \sin \frac{k_c x_1}{ r_0},
\end{aligned} \right.
 \end{equation}
 where
 \begin{equation}
 \label{eq5.11}
 \begin{aligned}
 & u_{k_c} = \frac{r_0 \pi}{k_c},   && v_{k_c} =1,\\
 & T_{k_c}=\frac{\alpha + \gamma (\Le \alpha_{k_c}^2 + \beta^1_{k_c})}{(\alpha_{k_c}^2 + \beta_{k_c}^1)(\Le \alpha_{k_c}^2 + \beta^1_{k_c})}  ,   && q_k=\frac{1}{\Le \alpha_{k_c}^2 + \beta^1_{k_c}}.
 \end{aligned}
 \end{equation}

Similarly, we can also derive from (\ref{eq5.6}) the conjugate eigenfunctions as follows

\begin{equation}
 \label{eq5.12}
 {\psi^1_{k_c}}^* = \left\{ \begin{aligned}
 & {u^1_1}^* = - u_{k_c}^* \cos \pi (x_2-r_0) \sin \frac{k_c x_1}{r_0},\\
  & {u^1_2}^* = v_{k_c}^* \sin \pi (x_2-r_0)\cos\frac{k_c x_1}{ r_0}, \\
  & {T^1_1}^* = T_{k_c}^* \sin \pi (x_2-r_0)\cos \frac{k_c x_1}{r_0},\\
  & {q^1}^* = q_{k_c}^* \sin \pi (x_2-r_0) \cos \frac{k_c x_1}{r_0},
 \end{aligned}\right. \end{equation}

\begin{equation}
\label{eq5.13}
{\psi^2_{k_c}}^*=\left\{ \begin{aligned}
 & {u^2_1}^* = u_{k_c}^* \cos \pi (x_2-r_0) \cos \frac{k_c x_1}{ r_0},\\
 & {u_2^2}^* = v_{k_c}^* \sin \pi (x_2-r_0) \sin \frac{k_c x_1}{ r_0},\\
 & {T^2}^* = T_{k_c}^* \sin \pi (x_2-r_0) \sin \frac{k_c x_1}{ r_0},\\
& {q^2}^* = q_{k_c}^* \sin \pi (x_2-r_0) \sin \frac{k_c x_1}{ r_0},
\end{aligned} \right.
 \end{equation}
 where
 \begin{equation}
 \label{eq5.14}
 \begin{aligned}
 & u_{k_c}^* = \frac{r_0 \pi}{ k_c},   && v_{k_c}^* =1, \\
 & T_{k_c}^* =\frac{\Pr R}{\alpha_{k_c}^2+\beta^1_{k_c}},   && q^*_{k_c} = \frac{\alpha \Pr R  -\Pr \tilde{R}(\alpha_{k_c}^2 + \beta^1_{k_c})}{(\alpha_{k_c}^2 + \beta_{k_c}^1)(\Le \alpha_{k_c}^2 + \beta^1_{k_c})}.
 \end{aligned}
 \end{equation}

{\it Step 4: Computation of the center manifold function.} The nonlinear operator $G$ defined in (\ref{eq5.2}) is bilinear, which can be expressed as
\begin{equation}
\label{eq5.15}
G(\psi, \phi) =- P \left( \begin{matrix}
u_1 \frac{\partial v_1}{ \partial x_1}+u_2 \frac{\partial v_1}{ \partial x_2} + \frac{1}{r_0} u_1 v_2     \\
 u_1\frac{\partial v_2}{\partial x_1} + u_2 \frac{\partial v_2}{\partial x_2} - \frac{1}{r_0} u_1 v_1  \\
u_1 \frac{\partial \tilde{T}}{\partial x_1} + u_2 \frac{\partial \tilde{T}}{\partial x_2}\\
 u_1 \frac{\partial \tilde{q} }{\partial x_1} + u_2 \frac{\partial \tilde{q}}{\partial x_2}
\end{matrix} \right),
\end{equation}
where $\psi = (u_1, u_2, T, q)$ and $\phi= (v_1, v_2, \tilde{T}, \tilde{q})$.
Direct calculation shows
\begin{equation}
\label{eq5.16}
\begin{aligned}
G(\psi^1_k, \psi^1_k) = & -P\Big[  \big(\frac{r_0 \pi^2}{2k} \sin \frac{2k x_1}{r_0},
\frac{\pi}{2} \sin 2 \pi (x_2 -r_0) - \frac{r_0 \pi^2}{4 k^2} [1+ \cos 2 \pi (x_2 - r_0)], 0,0 \big)^t \\
&+\big(-\frac{\pi}{4k} \sin 2 \pi (x_2 -r_0) \sin \frac{2 k x_1}{ r_0}, \frac{r_0 \pi^2}{ 4 k^2} \cos 2 \pi (x_2 - r_0) \cos \frac{2 k x_1}{r_0}, 0,0\big)^t\\
& + \big(0, \frac{r_0 \pi^2}{4 k^2} \cos \frac{2k x_1}{ r_0}, \frac{T_k \pi}{ 2} \sin 2 \pi (x_2 - r_0), \frac{q_k \pi}{2} \sin 2 \pi (x_2-r_0) \big)^t
\Big].
\end{aligned}
\end{equation}
Since the first two terms on the right hand side of  (\ref{eq5.16}) are gradient fields, we obtain
\begin{equation}
\label{eq5.17}
G(\psi^1_k, \psi^1_k) = \big(0, -\frac{r_0 \pi^2}{4 k^2} \cos \frac{2k x_1}{ r_0}, -\frac{T_k \pi}{ 2} \sin 2 \pi (x_2 - r_0), -\frac{q_k \pi}{2} \sin 2 \pi (x_2-r_0) \big)^t.
\end{equation}
Similarly, we can derive
\begin{equation}
\label{eq5.18}
\begin{aligned}
G(\psi^1_k, \psi^2_k) & = \big( -\frac{r_0 \pi^2}{2k}\cos 2\pi(x_2 -r_0)+\frac{\pi}{4k} \sin 2 \pi (x_2 -r_0), -\frac{r_0 \pi^2}{4 k^2} \sin \frac{2 k x_1}{r_0}, 0,0  \big)^t,\\
G(\psi^2_k, \psi^1_k) & = \big( \frac{r_0 \pi^2}{2k} \cos 2 \pi (x_2-r_0)-\frac{\pi}{4 k}\sin 2 \pi (x_2 - r_0), -\frac{r_0 \pi^2}{4k^2} \sin \frac{2k x_1}{r_0},  0,0 \big)^t,\\
G(\psi^2_k, \psi^2_k)&=\big( 0, \frac{r_0 \pi^2}{ 4 k^2} \cos \frac{2 k x_1}{ r_0}, -\frac{T_k \pi}{2} \sin 2 \pi(x_2 - r_0), -\frac{q_k \pi}{2} \sin 2 \pi (x_2-r_0)  \big)^t,
\end{aligned}
\end{equation}
and
\begin{equation}
\label{eq5.19}
\begin{aligned}
G(\psi^1_k, {\psi^1_k}^*) & = \big(0, -\frac{r_0 \pi^2}{4 k^2} \cos \frac{2k x_1}{ r_0}, -\frac{{T_k}^* \pi}{ 2} \sin 2 \pi (x_2 - r_0), -\frac{{q_k}^* \pi}{2} \sin 2 \pi (x_2-r_0) \big)^t,\\
G(\psi^1_k, {\psi^2_k}^*) & = \big( -\frac{r_0 \pi^2}{2k}\cos 2\pi(x_2 -r_0)+\frac{\pi}{4k} \sin 2 \pi (x_2 -r_0), -\frac{r_0 \pi^2}{4 k^2} \sin \frac{2 k x_1}{r_0}, 0,0  \big)^t,\\
G(\psi^2_k, {\psi^1_k}^*) & = \big( \frac{r_0 \pi^2}{2k} \cos 2 \pi (x_2-r_0)-\frac{\pi}{4 k}\sin 2 \pi (x_2 - r_0),   -\frac{r_0 \pi^2}{4k^2} \sin \frac{2k x_1}{r_0},0,0 \big)^t,\\
G(\psi^2_k, {\psi^2_k}^*)&=\big( 0, \frac{r_0 \pi^2}{ 4 k^2} \cos \frac{2 k x_1}{ r_0}, -\frac{{T_k}^* \pi}{2} \sin 2 \pi(x_2 - r_0), -\frac{{q_k}^* \pi}{2} \sin 2 \pi (x_2-r_0)  \big)^t,
\end{aligned}
\end{equation}
By (\ref{eq5.17}) - (\ref{eq5.18}), we have
\begin{equation}
\label{eq5.20}
<G(\psi^{i_1}_k, \psi^{i_2}_k), {\psi^{i_3}_k}^*>=0,
\end{equation}
for $i_1, i_2, i_3 =1,2$.
Moreover, direct calculation shows
\begin{equation}
\label{eq5.21}
\begin{aligned}
& <G(\phi_1, \phi_2), \phi_3  > = - <G(\phi_1, \phi_3), \phi_2>,\\
& <G(\phi_1, \psi^i_k),  {\psi^i_k}^*>=0,
\end{aligned}
\end{equation}
for $\phi_1, \phi_2, \phi_3 \in H_1$ and $i=1,2$.
Since the center manifold function $\Phi (x, y)= O(|x|^2+|y|^2)$, by (\ref{eq5.17})-(\ref{eq5.21}), we obtain
\begin{equation}
\label{eq5.22}
\begin{aligned}
<G(\psi), {\psi^1_{k_c}}^*> =&-x<G(\psi^1_{k_c}, {\psi^1_{k_c}}^* ), \Phi>-y<G(\psi^2_{k_c}, {\psi^1_{k_c}}^*), \Phi> \\
& + y<G(\Phi, \psi^2_{k_c}), {\psi^1_{k_c}}^*>+o(x^2 + y^2),\\
<G(\psi), {\psi^2_{k_c}}^*>=&-x<G(\psi^1_{k_c}, {\psi^2_{k_c}}^* ), \Phi>-y<G(\psi^2_{k_c}, {\psi^2_{k_c}}^*), \Phi> \\
& + x<G(\Phi, \psi^1_{k_c}), {\psi^2_{k_c}}^*>+o(x^2 + y^2).
\end{aligned}
\end{equation}

Let the center manifold function be denoted by
\begin{equation}
\label{eq5.23}
\Phi = \sum_{\beta \ne \beta^i_{kj}} \Phi_{\beta^i_{kj}}(x, y) \psi_{kj}^i.
\end{equation}

By (\ref{eq5.16}) - (\ref{eq5.22}), only
\begin{equation}
\begin{aligned}
\psi^1_{02}=& (0,0,0, \sin 2 \pi (x_2 - r_0)),\\
\psi^2_{02}=& (0,0, \sin 2 \pi (x_2 - r_0), 0 ),\\
\psi^3_{02}=& (\cos 2 \pi (x_2 -r_0), 0,0,0)
\end{aligned}
\end{equation}
contribute to the third order terms in evaluation of  (\ref{eq5.22}).
Direct calculation shows that
\begin{equation}
\begin{aligned}
<G(\psi^1_{k_c}, \psi^1_{k_c}) , \psi^1_{02}>&= -\frac{q_{k_c}}{2}\pi^2 r_0,   && <G(\psi^1_{k_c}, \psi^2_{k_c}), \psi^1_{02}  >= 0,  \\
<G(\psi^2_{k_c}, \psi^1_{k_c}),\psi^1_{02} >&=  0,
&&<G(\psi^2_{k_c}, \psi^2_{k_c}),\psi^1_{02}>= -\frac{q_{k_c}}{2}\pi^2 r_0,\\
 <G(\psi^1_{k_c}, \psi^1_{k_c}) , \psi^2_{02}>&= -\frac{T_{k_c}}{2}\pi^2 r_0,   && <G(\psi^1_{k_c}, \psi^2_{k_c}), \psi^2_{02}  >= 0,  \\
<G(\psi^2_{k_c}, \psi^1_{k_c}),\psi^2_{02} >&=  0,
&&<G(\psi^2_{k_c}, \psi^2_{k_c}),\psi^2_{02}>= -\frac{T_{k_c}}{2}\pi^2 r_0,\\
<G(\psi^1_{k_c}, \psi^1_{k_c}) , \psi^3_{02}>&= 0,   && <G(\psi^1_{k_c}, \psi^2_{k_c}), \psi^3_{02}  >= -\frac{r_0^2 \pi^3}{2{k_c}},  \\
<G(\psi^2_{k_c}, \psi^1_{k_c}),\psi^3_{02} >&=\frac{r_0^2 \pi^3}{2{k_c}} ,
&&<G(\psi^2_{k_c}, \psi^2_{k_c}),\psi^3_{02}>= 0.
     \end{aligned}
\end{equation}
We notice that
\begin{equation}
\begin{aligned}
\beta^1_{02} & = -\Le \alpha_{02}^2=-4 \pi^2\Len, \\
\beta^2_{02} & =-\alpha_{02}^2=-4 \pi^2,\\
\beta^3_{02} & = - \Pr \alpha^2_{02} - \frac{2 \Pr}{r_0^2}=-4 \pi^2 \Pr - \frac{1}{r_0^2} \Pr.
\end{aligned}
\end{equation}
 By the approximation formula of center manifold functions, see \cite{b-book}, we get
 \begin{equation}
  \label{eq5.27}
  \begin{aligned}
 & \Phi_{\beta^1_{02}} (x,y)=-\frac{q_k}{8 \pi \Len} (x^2+ y^2) + o(x^2+ y^2),\\
  & \Phi_{\beta^2_{02}} (x,y)=-\frac{T_k}{8 \pi } (x^2+ y^2)+ o(x^2+ y^2),\\
   &  \Phi_{\beta^3_{02}} (x,y)=o(x^2+ y^2).
         \end{aligned}
  \end{equation}
Plugging  (\ref{eq5.23}) into (\ref{eq5.22}), by   (\ref{eq5.19}) and (\ref{eq5.27}), we obtain
\begin{equation}
\label{eq5.28}
\begin{aligned}
<G(\psi), {\psi^1_{k_c}}^*> =&-x<G(\psi^1_{k_c}, {\psi^1_{k_c}}^* ), \Phi>+o(|x|^3+|y|^3) \\
                                    =&-\frac{\pi r_0}{16}(\frac{q_{k_c} q_{k_c}^*}{\Len} + T_{k_c} T_{k_c}^*) x(x^2 + y^2)+o(|x|^3+|y|^3),\\
 <G(\psi), {\psi^2_{k_c}}^*>=&-y<G(\psi^2_{k_c}, {\psi^2_{k_c}}^*), \Phi> +o(|x|^3+|y|^3)\\
 =& -\frac{\pi r_0}{16}(\frac{q_{k_c} q_{k_c}^*}{\Len} + T_{k_c} T_{k_c}^*) y(x^2 + y^2)+o(|x|^3+|y|^3).
\end{aligned}
\end{equation}
By \eqref{eq5.11}, \eqref{eq5.14} and \eqref{eq5.28}, we evaluate \eqref{eq5.8} at $R=R_c$ and obtain
the reduction equation
\begin{equation}
\begin{aligned}
& \frac{dx}{dt} = - \frac{b}{8}x(x^2+y^2) + o(|x|^3 + |y|^3),\\
& \frac{dy}{dt} = -\frac{b}{8}y(x^2+y^2)+o(|x|^3+|y|^3),
\end{aligned}
\end{equation}
where $b$ is as defined as
 \begin{equation}
 \label{eq5.1}
  b=\frac{\Len^3 \alpha_{k_c}^2R_c + \Big(\frac{(1+\Len^2)\alpha_1 \alpha_T h^2}{\alpha_q \kappa_T} + \frac{\alpha_1 \alpha_T \alpha_{k_c}^2 \Len^3 h^2}{24 \alpha_q \kappa_T} + \alpha_{k_c}^2 \Big)\tilde{R}}{ \frac{\Len^3 \alpha_{k_c}^6}{\Pr}(1+\frac{r_0^2 \pi^2}{k_c^2}) + \Len^3 \alpha_{k_c}^2 R + \Len\Big( \frac{(1+\Len)\alpha_1 \alpha_T h^2}{\alpha_q \kappa_T} + \frac{\alpha_1 \alpha_T \alpha_{k_c}^2 \Len^2 h^2}{24 \alpha_q \kappa_T} + \alpha_{k_c}^2   \Big) \tilde{R}} \,\, ,
  \end{equation}
and $\alpha_K^2$ and $R_c$ are given by Remark~\ref{rem4.2}.
It is obvious that $b>0$. Hence, standard energy estimate gives Assertions (2).
Assertions (1) and (3) follows from Theorem 3.1. This completes the proof of Theorem 4.1.

\section{Convection Scales}
Under the same setting as (\ref{eq2.10}) - (\ref{eq2.12}), including the fluid frictions, we consider the following nondimensional equation:
\begin{equation}
\label{eq6.1}
\begin{aligned}
\frac{\partial u_1}{ \partial t} = & \Pr \Big( \Delta u_1 + \frac{2}{r_0} \frac{\partial u_2}{ \partial x_1} -\frac{1}{r_0^2} u_1 -\delta_0 u_1- \frac{\partial p}{\partial x_1}\Big) \\
&-\omega u_2 - (u \cdot \nabla) u_1 - \frac{1}{r_0}u_1 u_2,\\
\frac{u_2}{\partial t} =& \Pr \Big( \Delta u_2 - \frac{2}{r_0}\frac{\partial u_1}{ \partial x_1} - \frac{2}{r_0^2}u_2 + RT+\tilde{R}q -\delta_1 u_2- \frac{\partial p}{\partial x_2}  \Big) \\
& + \omega u_1 - (u \cdot \nabla) u_2 + \frac{1}{r_0} u_1^2,\\
\frac{\partial T}{\partial t}=& \Delta T + \gamma u_2 + \alpha q - (u\cdot \nabla) T,\\
\frac{\partial q}{\partial t} = & \Le \Delta q + u_2 - (u \cdot \nabla)q,\\
\text{\it div }  u & =0,
\end{aligned}
\end{equation}
where
\begin{equation}
\label{eq6.2}
\begin{aligned}
& \gamma = 1 + \frac{\alpha}{24},\\
& \delta_i = C_i h^4,\,\, i=0,1,\\
& C_0 = 3.78\times 10^{-1}\,\, m^{-2}\cdot s^{-1}, \\
& C_1 = 6.7 \times 10^4 \,\, m^{-2} \cdot s^{-1}.
\end{aligned}
\end{equation}
 The domain is $M = [0, 2 \pi r_0 ] \times (r_0, r_0+ 1)$, and the boundary conditions are given by
 \begin{equation}
 \label{eq6.3}
 \begin{aligned}
 & (u, T, q) (x_1 + 2 \pi r_0, x_2)=(u, T, q)(x_1, x_2),\\
 & u_2 =0, \, \frac{\partial u_1}{\partial x_2} =0, \, T=0, \, q=0, \, \text{at }
x_2= r_0,\, r_0+1.
 \end{aligned}
 \end{equation}
 The corresponding eigenvalue problem reads
 \begin{equation}
 \label{eq6.4}
 \begin{aligned}
 & \Pr\Big( \Delta u_1 + \frac{2}{r_0} \frac{\partial u_2}{ \partial x_1} - \frac{1}{r_0^2} u_1-\delta_0 u_1 - \frac{\partial p}{\partial x_1}  \Big) - \omega u_2 = \beta u_1,\\
 & \Pr \Big( \Delta u_2 - \frac{2}{r_0}\frac{\partial u_1}{\partial x_1}  - \frac{2}{r_0^2} u_2 + RT +\tilde{R} q -\delta_1 u_2- \frac{ \partial p}{\partial x_2} \Big)+ \omega u_1 = \beta u_2, \\
 & \Delta T + \gamma u_2+  \alpha q = \beta T,\\
 & \Le \Delta q + u_2 = \beta q,\\
 & \frac{\partial u_1}{ \partial x_1} + \frac{\partial u_2}{\partial x_2} = 0,
 \end{aligned}
 \end{equation}
 Using the same analysis as in Section 3.1, we
 obtain the following formula of the critical thermal Rayleigh number
 \begin{equation}
\label{eq6.5}
\begin{aligned}
R_c = & \min_{k,j \ge 1} \Big[ -\Big(\frac{1}{\Len}+(1+\frac{\Le \alpha_{kj}^2}{24}) \frac{\alpha_1 \alpha_T h^2}{\Le \alpha_q \alpha_{kj}^2 \kappa_T}    \Big) \tilde{R}  \\
& + \alpha^4_{kj}\frac{r_0^2}{k^2} (\alpha^2_{kj}+\frac{1}{r_0^2})+\frac{\alpha_{kj}^2}{r_0^2}
+\delta_1 \alpha_{kj}^2 +\frac{\delta_0 \pi^2 r_0^2 \alpha_{kj}^2}{k^2}\Big]\\
=& \min_{k \ge 1} \Big[ -\Big(\frac{1}{\Len}+(1+\frac{\Le \alpha_{k1}^2}{24}) \frac{\alpha_1 \alpha_T h^2}{\Le \alpha_q \alpha_{k1}^2 \kappa_T}    \Big) \tilde{R} \\
& + \alpha^4_{k1}\frac{r_0^2}{k^2} (\alpha^2_{k1}+\frac{1}{r_0^2})+\frac{\alpha_{k1}^2}{r_0^2}
 +\delta_1 \alpha_{k1}^2 +\frac{\delta_0 \pi^2 r_0^2 \alpha_{k1}^2}{k^2}
 \Big],
\end{aligned}
\end{equation}
where
$$
\alpha_{k1}^2= \pi^2 + \frac{k^2}{r_0^2}.
$$
Let $y=\frac{k^2}{r_0^2}$ and
\begin{equation*}
\begin{aligned}
g(y)=& -\Big(\frac{1}{\Len}+(1+\frac{\Len(y+\pi^2)}{24}  )\frac{\alpha_1 \alpha_T h^2}{\Le \alpha_q \kappa_T(y+\pi^2) }  \Big)\tilde{R} + \frac{1}{y}(y+\pi^2)^2(y+\pi^2+\frac{1}{r_0^2}) \\
&+ \frac{1}{r_0^2}(y+\pi^2) + \delta_1(y+\pi^2)+\frac{\delta_0 \pi^2 (y+\pi^2)}{y}.
\end{aligned}
\end{equation*}
Taking the derivative of $g(y)$, we get
\begin{equation*}
g'(y)= 2y+(3 \pi^2 +\frac{2}{r_0^2} + \delta_1) - \frac{1}{y^2}(\pi^6 + \frac{\pi^4}{r_0^2} + \delta_0 \pi^4) + \frac{\alpha_1 \alpha_T h^2 \tilde{R}}{\Le \alpha_q \kappa_T (y+\pi^2)^2 }.
\end{equation*}
By (\ref{eq6.2}), we have
\begin{equation}
\label{eq6.7}
\delta_1 = 2.76 \times 10^{20}, \qquad \delta_0 = 1.55 \times 10^{15}.
\end{equation}
Hence,
\begin{equation}
\label{eq6.6}
\delta_1 \gg \delta_0 \gg 1.
\end{equation}

Under this condition, the critical value of $g(y)$ is approximated by
\begin{equation}
\label{eq6.8}
y_c \simeq \Big( \frac{\delta_0}{\delta_1}\Big) \pi^2.
\end{equation}
Therefore, the critical thermal Rayleigh number is approximated by
\begin{equation}
\label{eq6.9}
R_c \simeq \delta_1 \pi^2=2.76 \times 10^{21}.
\end{equation}
Next, as in \cite{ptd, MW3}, we adopt
\begin{equation}
\begin{aligned}
& \nu = 1.6 \times 10^{-5}\,\, m^2/s,  && \kappa_{T} = 2.25 \times 10^{-5} \,\, m^2/s,\\
& \alpha_{T}=3.3 \times 10^{-3} /^{\circ}C,    && \Pr = 0.71.
\end{aligned}
\end{equation}
We notice also that the height of the troposphere is
$$
h = 8 \times 10^3 \,\, m.
$$
Hence, we get
\begin{equation}
T_0 -T_1 = \Delta T_c = \frac{\kappa_T \nu}{\alpha_{T} g h^3} R_c = 60^{\circ} C.
\end{equation}

Here we notice that the above approximations agree with that of the model of  tropical atmospheric circulations without humidity. However, as we can see from (\ref{eq6.5}), the coefficient of $\tilde{R}$ is negative. This implies that the humidity factor lowers down the critical thermal Rayleigh number a little bit.

Next, since the nodimensional radius of earth is $r_0 = 6400000/ h =800$, we derive from 
\begin{equation}
 \frac{k^2}{r_0^2} = \Big( \frac{\delta_0}{\delta_1}\Big)^{1/2} \pi^2
\end{equation}
that the wave number $k_c$ and the convection length scale $L_c$  as
$$
k_c \simeq 6, \qquad L= (6400 \times \pi) / 6 = 3350 \,\, km.
$$
This is consistent with the large scale atmospheric circulation over the tropics.

\end{document}